\documentclass[preprint]{aastex61}

\received{May 31, 2018}
\revised{September 12, 2018}
\accepted{November 13, 2018}

\shorttitle{giant impact on an ice giant}
\shortauthors{Kurosaki \& Inutsuka}

\begin{document}

\title{The Exchange of Mass and Angular Momentum in the Impact Event of Ice Giant Planets: Implications for the origin of Uranus}

\correspondingauthor{Kenji Kurosaki}
\email{kurosaki.k@nagoya-u.jp}

\author{Kenji Kurosaki}
\affil{Department of Physics, Nagoya University, Chikusa-ku, Nagoya 464-8602, Japan}

\author{Shu-ichiro Inutsuka}
\affil{Department of Physics, Nagoya University, Chikusa-ku, Nagoya 464-8602, Japan}

\begin{abstract}

Uranus has a tilted rotation axis, which is supposed to be caused by a giant impact. In general, an impact event also changes the internal compositional distribution and drives mass ejection from the planet, which may provide the origin of satellites. Previous studies of the impact simulation of Uranus investigated the resultant angular momentum and the ejected mass distribution. However, the effect of changing the initial condition of the thermal and compositional structure is not studied. In this paper, we perform hydrodynamics simulations for the impact events of Uranus-size ice giants composed of a water core surrounded by a hydrogen envelope using two variant methods of the smoothed particle hydrodynamics.
We find that the higher entropy target loses its envelope more efficiently than the low entropy target. However, the higher entropy target gains more angular momentum than the lower entropy target since the higher entropy target has more expanded envelope.
We discuss the efficiency of angular momentum transport and the amount of the ejected mass and find a simple analytical model to roughly reproduce the outcomes of numerical simulations. We suggest the range of possible initial conditions for the giant impact on proto-Uranus that reproduces the present rotation tilt of Uranus and sufficiently provides the total angular momentum of the satellite system that can be created from the fragments from the giant impact. 

\end{abstract}


\section{Introduction} \label{sec:intro}
Our solar system has two ice giants, Uranus and Neptune,
that are supposed to be mainly composed of gas envelope, icy mantle and solid core from the top to the bottom \citep[e.g.][]{Hubbard1980, Helled2011, Nettelmann2016}.
These two planets have similar mass and radius,
while their spin axes, satellite systems, and intrinsic luminosities are different.
For example, the obliquity of Uranus is 98$^\circ$, while that of Neptune is 27.7$^\circ$.
\citet{Safronov1966} pointed out that Uranus experienced a giant impact event to reproduce its tilted rotation axis.
An impactor of several earth mass may have
transported the angular momentum to proto-Uranus via collision and 
tilted the rotation axis of proto-Uranus.
\citet{Parisi1997} estimated the giant impact based on the conservation of angular momentum and energy and concluded
that the minimum impactor mass is $\sim 1-1.1 M_\oplus$.
Such large impact event may also have produce a circumplanetary disk around the proto-Uranus, which might be the origin of small prograde satellites around Uranus \citep[e.g.,][]{Parisi2008}.
A giant impact scenario is widely accepted as an explanation for terrestrial moon formation \citep{Hartmann1975,Cameron1976} and recently suggested to apply to Phobos and Deimos formation 
\citep[e.g.,][]{Citron2015,Hyodo2017a}.
Since the impact event causes erosion of proto-Uranus, the angular momentum that impactor brings
is redistributed to proto-Uranus and eroded gas envelope and fragments around the proto-Uranus.
The internal compositional distribution after the impact is unknown because the mixing process by the impact event is a 
highly non-liner problem to analyze theoretically. 
Moreover, the impact event is also essential to the thermal evolution. 
Present Uranus has a low intrinsic luminosity comparing to Neptune.
\citet{Nettelmann2016} explained the luminosity of Uranus considering a thermal boundary layer between an outer H-He-rich envelope and an inner ice-rich layer.
If a giant impact event occurs, internal compositional distribution should be changed.
If the mixing of icy material in the H$_2$/He envelope is efficient, the thermal evolution is expected to be accelerated \citep{Kurosaki2017}.

The giant impact simulations on rock-composed protoplanets have been systematically calculated \citep{Marcus2009,Genda2012}.
Those studies have not considered the atmosphere.
The atmosphere is eroded more efficiently than rock materials because the thermal pressure of gas is more sensitive than that of rock materials. 
Impact events on Uranus have been investigated by use of hydrodynamical simulation.
\citet{Korycansky1990} studied a giant impact on ice giant for the first time.
They calculated one-dimensional spherically symmetric hydrodynamic simulation
and investigated the H$_2$/He envelope erosion due to the impact.
They found a sharp transition between the cases of nearly complete retention and dispersal of  H$_2$/He envelope, which depend on the amount of energy deposition on the envelope.
Three-dimensional hydrodynamical simulations are done by \citet{Slattery1992} and \citet{Kegerreis2018}.
\citet{Slattery1992} used the smoothed particle hydrodynamic simulation (hereafter SPH simulation) 
to constrain the angular momentum of proto-Uranus and eroded mass after the giant impact.

Previous work constrained the impactor mass to explain the amount of the present angular momentum of Uranus.
In the context of thermal evolution of Uranus, the evolutionary stage at time of the giant impact is also important. 
When the age of  proto-Uranus is younger than 10$^8$~years, 
the H$_2$/He envelope still remains extended and the efficiency of the gas envelope erosion will be increased.
It should be useful to study the effect of changing the structure of the proto-Uranus on the result of giant impact to constrain the evolutionary stage of the proto-Uranus at the time of the giant impact event. 

In this paper, we study the impact event on a proto-Uranus.
The aim of this paper is to investigate the envelope erosion and the efficiency of angular momentum transport to the proto-Uranus.
Since there are no constraints on the age of Uranus at the time of the impact event, we consider two extreme cases: an impact onto a young ($10^8$ years) ice giant, when it was in a high-temperature state, and an impact onto a mature ice giant ($10^9$ years), when it was in a low-temperature state.

Section~\ref{method} describes the methods and settings for our simulation.
The results of our study are described in Section~\ref{results} and discussions in Section~\ref{discussion}.
We summarize the conclusion of our study in section~\ref{conclusion}.

\section{Method}~\label{method}
We use Godunov-type smoothed particle hydrodynamical calculation, hereafter GSPH,
\citep{Inutsuka2002,Sugiura2016} to solve the following hydrodynamic equations:
\begin{eqnarray}
\frac{d\rho}{dt} &=& -\rho \nabla\cdot\mathbf{v} \\
\frac{d\mathbf{v}}{dt} &=& -\frac{1}{\rho} \nabla P + \nabla \int dx'^3 \frac{G\rho(x')}{|\mathbf{x}-\mathbf{x'}|} \\
\frac{du}{dt} &=& -\frac{P}{\rho} \nabla \cdot\mathbf{v} \\
P &=& P(\rho, u)  \label{eos}
\end{eqnarray}
where $\rho, P, \mathbf{v}$ and $u$ are density, pressure, velocity, and specific internal energy, respectively.
$t$ is the time, $\mathbf{x}$ is the position, and
$G(=6.67408\times10^{-8}~\mathrm{cm}^3~\mathrm{g}^{-1}~\mathrm{s}^{-2})$ is the gravitational constant.
The method has advantages on tracing strong shock and contact discontinuity \citep{Cha2010}.
We introduce the exact and spatially second-order Riemann solver for piece-wise polytropic gas to our GSPH.
As for the equation of state (Eq.~\ref{eos}), we use \citet{Saumon1995} for hydrogen and helium, and SESAME 7150 for water \citep{SESAME}.
To implement non-ideal equation of states in GSPH, 
the effective heat ratio $\gamma_\mathrm{eff}$  is calculated from the data table of non-ideal equation of state. 
We also take into account the gravity force on Riemann solver \citep[submitted]{Guo}.
We have implemented the acceleration modules for our SPH code with FDPS \citep{Iwasawa2016} and FDPS fortran interface \citep{Namekata2018}. 
We have tested our SPH code by reproducing the analytical solution for a shock tube and the Lane-Emden solution for a polytrope gas equilibrium spheres.

\subsection{Initial conditions for target and impactor}
In this paper, we fix following two parameters.
We assume masses of target and impactor are 13~$M_\oplus$ and 1~$M_\oplus$, respectively.
The target is composed of 20~\% of hydrogen and 80~\% of water. 
The impactor is composed of 100~\% of water. 
The impact angle, hereafter $\theta$, is assumed to $15^\circ, 30^\circ, 45^\circ,$ and $60^\circ$.
The impact position is $(R_p+r_\mathrm{imp})\sin\theta$, where $R_p$ and $r_\mathrm{imp}$ are radii of target and impactor, respectively.
The impact velocity is assumed to be the escape velocity, which is represented by $v_\mathrm{imp} = \sqrt{2GM_p/R_p}$
\footnote{Based on the two-body problem, the impact velocity is equal to $v_\mathrm{imp,2} = \sqrt{2G(M_p+m_\mathrm{imp})/(R_p+r_\mathrm{imp})}$. The adopted impact velocity ($v_\mathrm{imp})$ is faster than $v_\mathrm{imp,2}$ by 10~\%.}
where $M_p$ is the target mass, and $m_\mathrm{imp}$ is the impactor mass.
The impact velocities are $1.68 \times 10^{4}$~m~s$^{-1}$ and $1.85\times 10^{4}$~m~s$^{-1}$ for the HT and LT target discussed below, respectively.
Those impact velocities correspond to the free-fall motion from infinity.
The escape velocity of the target depends on the target's radius. 
Since the HT target has a larger radius than the LT target, its escape velocity is lower.
Here we assume that both the target and impactor are not spinning before the collision,
as in \citet{Slattery1992} and \citet{Kegerreis2018}.
We also investigate the effect of changing temperature structure of the target on the outcome of the giant impact.
Here we assume two types of target.
The high temperature target, hereafter HT, is assumed to have entropy $S=S(100~\mathrm{bar},1000~\mathrm{K})$, while the low temperature target , hereafter LT, is assumed its entropy $S=S(100~\mathrm{bar},500~\mathrm{K})$.
Table~\ref{param_imp_target} shows the parameters for the target and impactor.
Both the HT and the LT target are warmer than present Uranus.
The age of LT target is 10$^9$ years, while HT target is 10$^8$ years (see Appendix~B). The target \citet{Kegerreis2018} adopted is equivalent to the present Uranus.
If the rotation period of the target is longer than 
$\sim 100$ hours, the target's angular momentum is smaller than the present Uranus by an order of magnitude and the initial target's spin can be ignored.
If the target's angular momentum is comparable to the present Uranus, the rotating H-He atmosphere changes its impact mach number and hence the resulting internal angular momentum distribution.
However, it is beyond the scope of this study.
We assume that the target's angular momentum is much smaller than that provided by impactor for simplicity.
Hereafter we discuss the spin of the target that is the result of the impact.

We stop the numerical simulation at $t=10~t_\mathrm{ff}$, where $t_\mathrm{ff}$ is the free fall time of the target given by $t_\mathrm{ff}=\sqrt{\pi^2 R_p^3/(8GM_p)}$.
$t_\mathrm{ff}$ for HT and LT are 0.95 hours and 0.71 hours, respectively.
Thus we stop HT and LT simulations at 9.5 hours and 7.1 hours, respectively.
In this case, the number of timesteps in our simulations are only on the order of $10^3$ thanks to the efficiency of GSPH.

\begin{table}[h]
\begin{center}
\begin{tabular}{c|c|c|c|c|c|c}
& Mass [g] & Radius [cm] & Temperature [K] & H$_2$ [wt~\%] & H$_2$O [wt~\%] & Particles \\
\hline
HT & $7.82\times10^{28}$ & $3.18\times10^{9}$ & 1000~K (100~bar) & 20 & 80 & 65500 \\
\hline
LT & $7.82\times10^{28}$ & $3.03\times10^{9}$ & 500~K (100~bar) & 20 & 80 & 65500 \\
\hline
Impactor & $5.97\times10^{27}$ & $1.13\times10^{9}$ & 300~K (1~bar) & 0 & 100 & 5000
\end{tabular}
\caption{\label{param_imp_target} Conditions for HT target, LT target, and impactor for our simulation.}
\end{center}
\end{table}

\section{Results}~\label{results}
\begin{figure}[h]
\begin{center}
\includegraphics[width=14cm, bb=10 10 400 400]{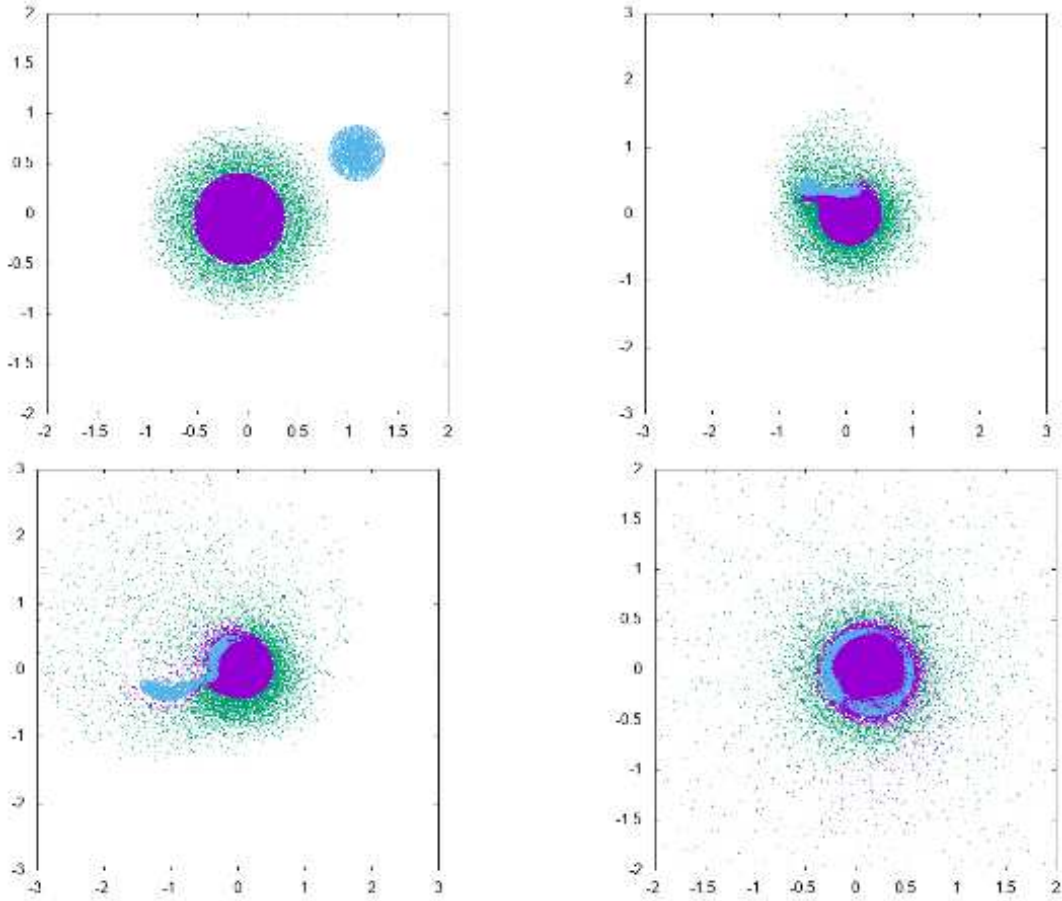}
\caption{\label{fig_timelapse} Snapshots of the high temperature target's impact simulation whose impact angle is $\theta=30^\circ$. Green, purple, and blue dots represent target's hydrogen particles, target's water particles, and impactor's water particles, respectively. Those snapshots represent initial condition $(t=0)$, $t=t_\mathrm{ff}$, $t=2~t_\mathrm{ff}$, and $t=8~t_\mathrm{ff}$ from top-left, top-right, bottom-left, and bottom-right, respectively, where $t_\mathrm{ff}$ is a free fall time.}
\end{center}
\end{figure}

In this section, we show results of our impact simulations.
Figure~\ref{fig_timelapse} shows four snapshots of the impact simulation for HT whose impact angle is $30^\circ$, which represent the initial condition $(t=0)$, $t=t_\mathrm{ff}$, $t=2~t_\mathrm{ff}$, and $t=8~t_\mathrm{ff}$.
When the impactor collides on the target, 
the impactor gives its momentum to the hydrogen envelope of the target.
In the case of $\theta=30^\circ$,
the impactor collide with the H$_2$O core of the target.
After the collision, the impactor fall onto the target's core and the hydrogen envelope is eroded due to the collision.  

Here we introduce the definition for the eroded particle from the target at $t=10~t_\mathrm{ff}$ after the collision.
The mass, position, velocity, and internal energy of $i-$th particle are $m_i, \mathbf{r}_i$, $\mathbf{v}_i$ and $u_i$, respectively.
After the impact, the hydrogen gas expands around the target.
Eroded gas particles have enough energy to escape from the target and their positions are outside of the target.
In this study, the eroded particle condition is
\begin{eqnarray}
\frac{1}{2}m_i |\mathbf{v}_i|^2 - m_i \sum_{i\ne j}\frac{Gm_j}{|\mathbf{r}_i - \mathbf{r}_j|}  &>& 0 \label{lost_cond_1} \\
|\mathbf{r}_i - \mathbf{R}_{t, c} | &>& R_p \label{lost_cond_2}
\end{eqnarray}
where $\mathbf{R}_{t, c}$ is the position of the center of the target.

\begin{figure}[h]
\begin{center}
\includegraphics[width=14cm, bb=10 10 400 400]{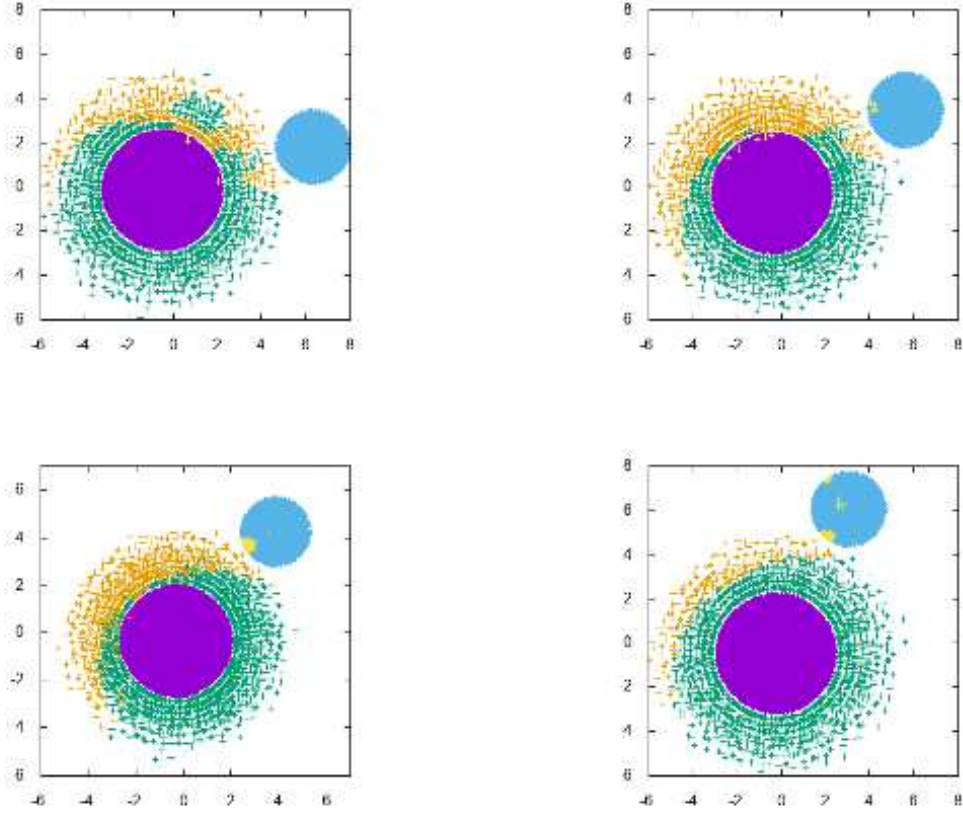}
\caption{\label{fig_eject_region_HT} Ejected region for high temperature target. Only the particles close to the midplane $(-0.1 < z < 0.1)$ are shown. Green, purple, and blue dots represent target's hydrogen particles, target's water particles, and impactor's water particles, respectively. Orange and yellow dots shows eroded hydrogen and water particles, respectively.
We represent different impact angles as follows. Top-left, top-right, bottom-left, and bottom-right represent 15$^\circ$, 30$^\circ$, 45$^\circ$, and 60$^\circ$, respectively. Those figures show a superposition of the initial state and the state after the collision.
}
\end{center}
\end{figure}

\begin{figure}[h]
\begin{center}
\includegraphics[width=14cm, bb=10 10 400 400]{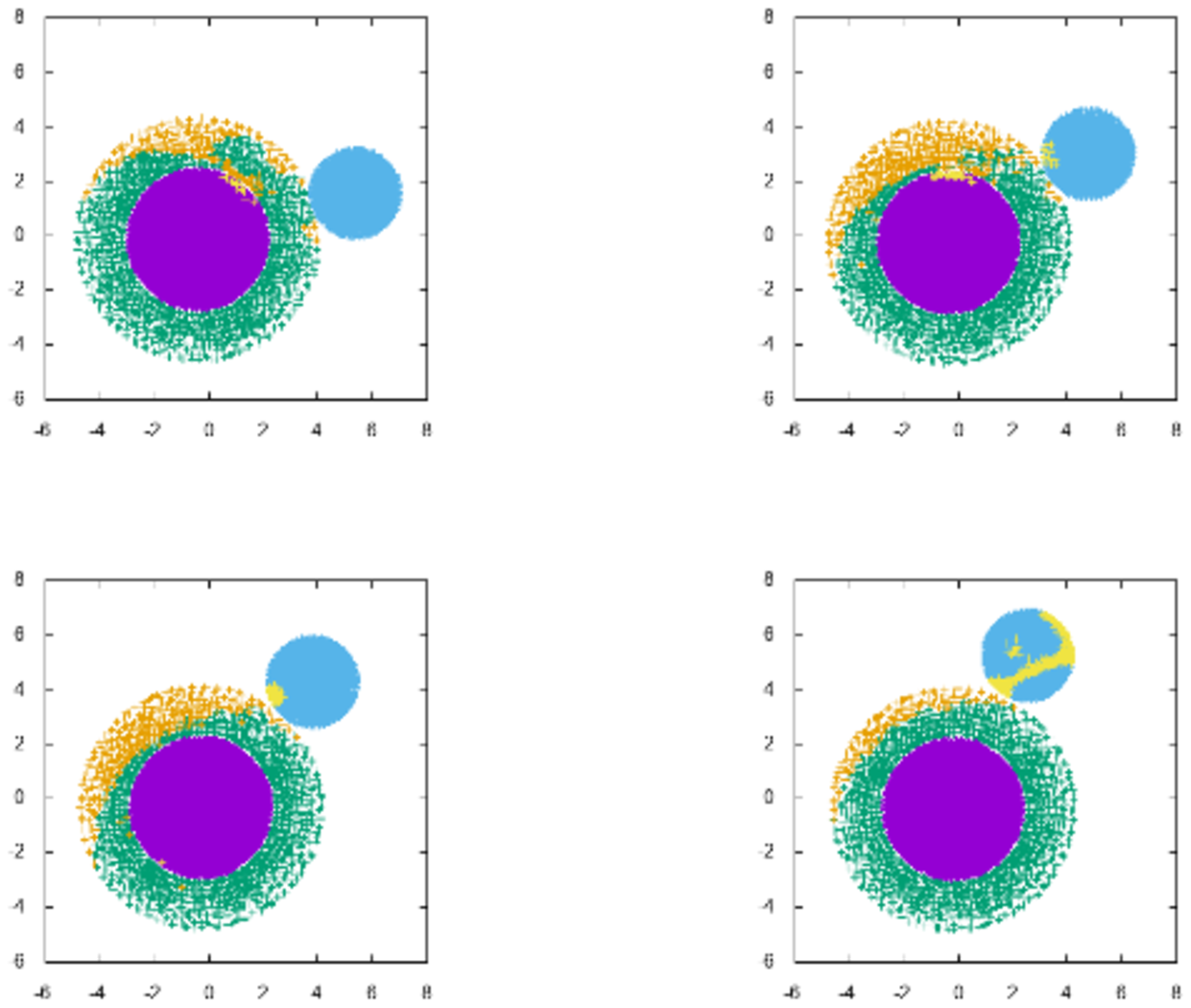}
\caption{\label{fig_eject_region_LT} Same as Figure \ref{fig_eject_region_HT} but for LT target.}
\end{center}
\end{figure}

Figure~\ref{fig_eject_region_HT} and \ref{fig_eject_region_LT} show the eroded region after the collision for the high temperature target and low temperature target, respectively.
The eroded region for hydrogen and water is shown in orange and yellow, respectively.
Those figures show the z-plane cross sections of the results by choosing the particles in the range of $z=[-0.1,0.1]$. 
We can find that the property of the mass erosion changes depending on whether the impactor collides with the target core or not.
When the impactor collides with the core, 
the impactor change its trajectory and the hydrogen envelope is eroded along the impactor's trajectory.
In the case of impact angle is $15^\circ$ and $30^\circ$,
the impactor collide with the water core of the target.
On the other hand, the impactor does not collide with the water core in the case of impact angle is $45^\circ$ and $60^\circ$.
After the collision, 
the impactor is captured in the target 
while part of the hydrogen envelope of the target
receives energy and angular momentum from the impactor and escapes from the target.

Figure~\ref{fig_mloss} shows the relationship between the eroded mass and impact parameter for HT and LT case.
In HT case, the eroded mass is larger than LT case.
That is because the volume of hydrogen envelope of HT case is larger than that of LT case and the escaped region of HT case is also larger than LT case.
Moreover, the escape velocity at the surface of HT is smaller than LT, which promotes the ejection of particles from the target.
After the collision, the minimum masses of H-He retained for HT and LT are 80~\% and 90~\%, respectively, while \citet{Kegerreis2018} obtain 75~\%.
The difference is due to the equation of states for hydrogen-helium.
The adopted equations of states for hydrogen helium \citep{Saumon1995} in our calculation have smaller heat capacity ratio compared to that of \citet{Hubbard1980}.
That is, in our calculation, the pressure response against density is softer than previous study. 
We think this is the reason for the difference between our results and that of \citet{Kegerreis2018}.
The eroded mass of ice for small impact angle is smaller than that of hydrogen by an order of magnitude. In the case of low impact angle, both of HT and LT lose little water from core and impactor. However, the eroded water mass is more than the total mass of Uranian satellites. In the case of high impact angle, the impactor escapes from the target and the mass loss of water increases. 

\section{Discussion} \label{discussion}
\begin{figure}[h]
\begin{center}
\includegraphics[width=7cm]{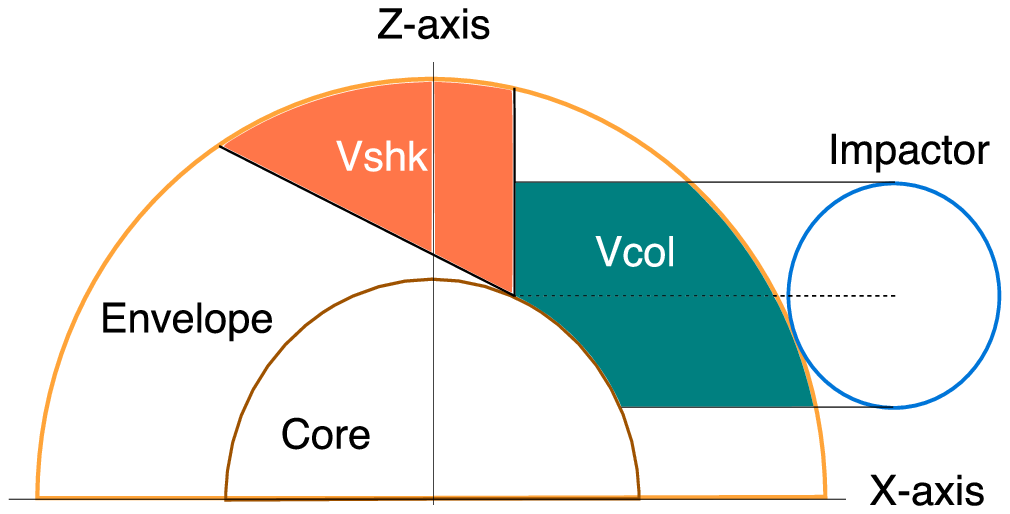}
\includegraphics[width=7cm]{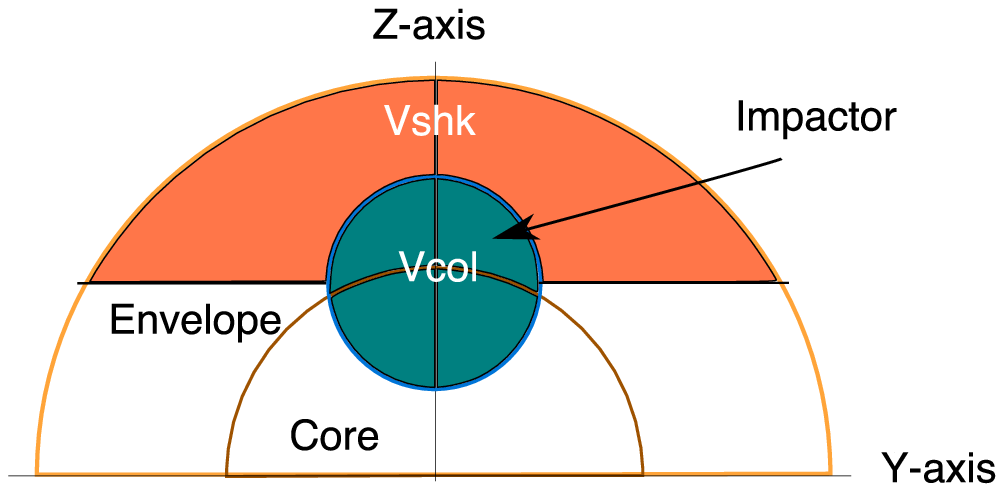}
\caption{\label{fig_mloss_anal}
Panels show the estimated eroded region.
The left and right panels shows the $V_\mathrm{col}$ and $V_\mathrm{shk}$ on XZ-plane and YZ-plane, respectively.
The green region shows $V_\mathrm{col}$ estimated by Eq.~\ref{vcol1}.
The orange region shows $V_\mathrm{shk}$ estimated by Eq.~\ref{vcol2}.}
\end{center}
\end{figure}

\begin{figure}[h]
\begin{center}
\includegraphics[width=7cm]{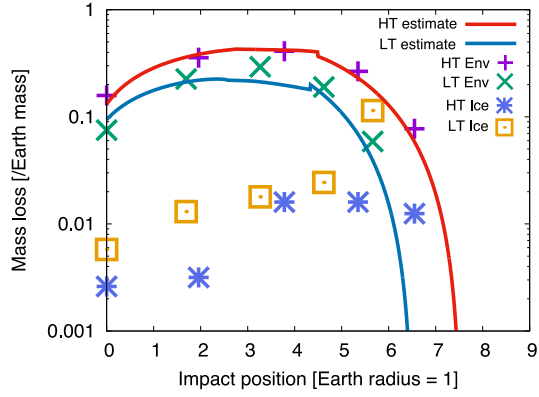}
\caption{\label{fig_mloss}
Relationship between the eroded mass and the impact parameter.
The purple and green symbols represents the eroded hydrogen mass from the HT and LT, respectively. The dark blue and orange symbols represents the eroded water mass of HT+impactor and LT+impactor, respectively.
The red and blue lines represent the analytical solution for the high temperature target and low temperature target derived by Eq.~\ref{mesc_target}.
}
\end{center}
\end{figure}

The eroded mass due to the collision can be understood by considering the heating area.
Here we estimate the volume of ejected mass in terms of collision induced erosion volume $V_\mathrm{col}$ and shock induced erosion volume, $V_\mathrm{shk}$, as
\begin{equation}
V_\mathrm{ej} =  V_\mathrm{col} + V_\mathrm{shk}
\end{equation}
where
\begin{eqnarray}
V_\mathrm{col} & = & \sigma (R_p-R_c)\cos\theta \label{vcol1} \\
V_\mathrm{shk} & = &  \left[ \frac{\pi}{6} h_1^2(3R_p-h_1) - \frac{\pi}{6} h_2^2(3R_c-h_2) \right] \\
&& - \left[ \frac{\pi}{6} h_3^2(3R_p-h_3) - \frac{2\pi R_c^3}{3} - \frac{4\pi }{3} (R_p^3-R_c^3) \frac{\varphi}{2 \pi} \right]  \label{vcol2}
\end{eqnarray}
where $h_1=R_p+R_c\cos\varphi$, $h_2=R_c+R_c\cos\varphi$, $h_3=R_p+R_c$, $\varphi = \sin^{-1}[(R_p+r_\mathrm{imp})\sin\theta / R_c]$, $R_c$ is the core radius of the target, and $r_\mathrm{imp}$ is the radius of the impactor (see Fig.~\ref{fig_mloss_anal}).
Thus, we can estimate the escaped mass as
\begin{equation}
M_\mathrm{esc, t} = \overline{\rho_\mathrm{env}} V_\mathrm{area} \label{mesc_target}
\end{equation}
where $\overline{\rho_\mathrm{env}}$ is the averaged density.
Since there is a strong negative gradient of density in hydrogen envelope, we adopt the bottom density of hydrogen envelope as the characteristic density.
The density of hydrogen envelope $\approx 0.1 \mathrm{g}\cdot\mathrm{cm}^{-3}$ at the core-envelope boundary for HT and LT.
Here we assume $\overline{\rho_\mathrm{env}}=0.1~\mathrm{g}~\mathrm{cm}^{-3}$.
When the impact parameter is large, a part of the impactor does not collide with the envelope of the target
and then it does not fall onto the target.
The impactor's loss $m_\mathrm{esc, imp}$ can be estimated as
\begin{equation}
m_\mathrm{esc, imp} = \overline{\rho_\mathrm{imp}}
\pi r_\mathrm{imp}^3 \left[ \frac{2}{3} + \frac{(R_p+r_\mathrm{imp})\sin\theta - R_p}{r_\mathrm{imp}} - \frac{1}{3} \left( \frac{(R_p+r_\mathrm{imp})\sin\theta - R_p}{r_\mathrm{imp}} \right)^3 \right] \label{mesc_impactor}
\end{equation}
for $\sin\theta \ge \frac{R_p-r_\mathrm{imp}}{R_p+r_\mathrm{imp}}$.
If $\sin\theta < \frac{R_p-r_\mathrm{imp}}{R_p+r_\mathrm{imp}}$,
the entire impactor is expected to fall onto the target.
That is, we set $m_\mathrm{esc, imp} = 0$.
Figure~\ref{fig_mloss} also shows the analytical result.
The analytical model can reproduce the trend of the relationship between the eroded mass and the impact parameter.
However, the analytical model does not include the effect of the escape velocity, which should determine the difference between the HT and LT quantitatively.
We suggest that it will be important to determine the propagation of three-dimensional shock wave in the interior of HT and LT.

The total angular momentum that is transferred from the impactor to the target via collision is $L_\mathrm{tot}(=m_\mathrm{imp} v_\mathrm{imp} (R_p+r_\mathrm{imp}) \sin\theta)$.
After the collision, 
the eroded mass and 
remove some fraction of the angular momentum that was given by the impactor.
The angular momentum that is transferred 
thus
\begin{eqnarray}
L_p & = & L_\mathrm{+} - M_\mathrm{esc, t} v_\mathrm{ej} R_c - m_\mathrm{esc, imp} v_\mathrm{imp} (R_p+r_\mathrm{imp}) \sin\theta \label{Lp_target}
\label{Lp_target_bimp}
\end{eqnarray}
where
$L_+$ is transferred angular momentum to the target by the impactor
and $v_\mathrm{ej}$ the ejecta velocity.
Here we estimate $v_\mathrm{ej}$ by the impact ejecta scaling law \citep{Melosh1989,Richardson2005}.
\begin{equation}
v_\mathrm{ej} = \frac{2 \sqrt{R_\mathrm{c}g_s}}{1+\varepsilon} \label{vej_melosh}
\end{equation}
where $g_s$ is the surface gravity, and $\varepsilon$ is a material constant.
When all the impactor particle are exist in $\theta>0$, all particles moves counterclockwise after the impact that means $L_{+}$ is equal to $L_\mathrm{tot}$.
When $(R_p+r_\mathrm{imp})\sin\theta < r_\mathrm{imp}$ is satisfied, 
some particles whose positions are $\theta<0$ impact and reduce $L_{+}$.
Here we divide the impactor to $\theta>0$ and $\theta<0$ region.
$V_1, V_2$ are the volume of $\theta>0$ and $\theta<0$ region and $h_1, h_2$ are distance from the $x-z$ plane.
Thus, 
\begin{equation}
L_{+} = \rho_\mathrm{imp}v_\mathrm{imp} [ V_1 (r_\mathrm{imp} +(R_p+r_\mathrm{imp})\sin\theta - h_1 ) - V_2 (r-(R_p+r_\mathrm{imp})\sin\theta - h_2) ] \label{Lp_w_anti}.
\end{equation}

\begin{figure}[h]
\begin{center}
\includegraphics[width=7cm]{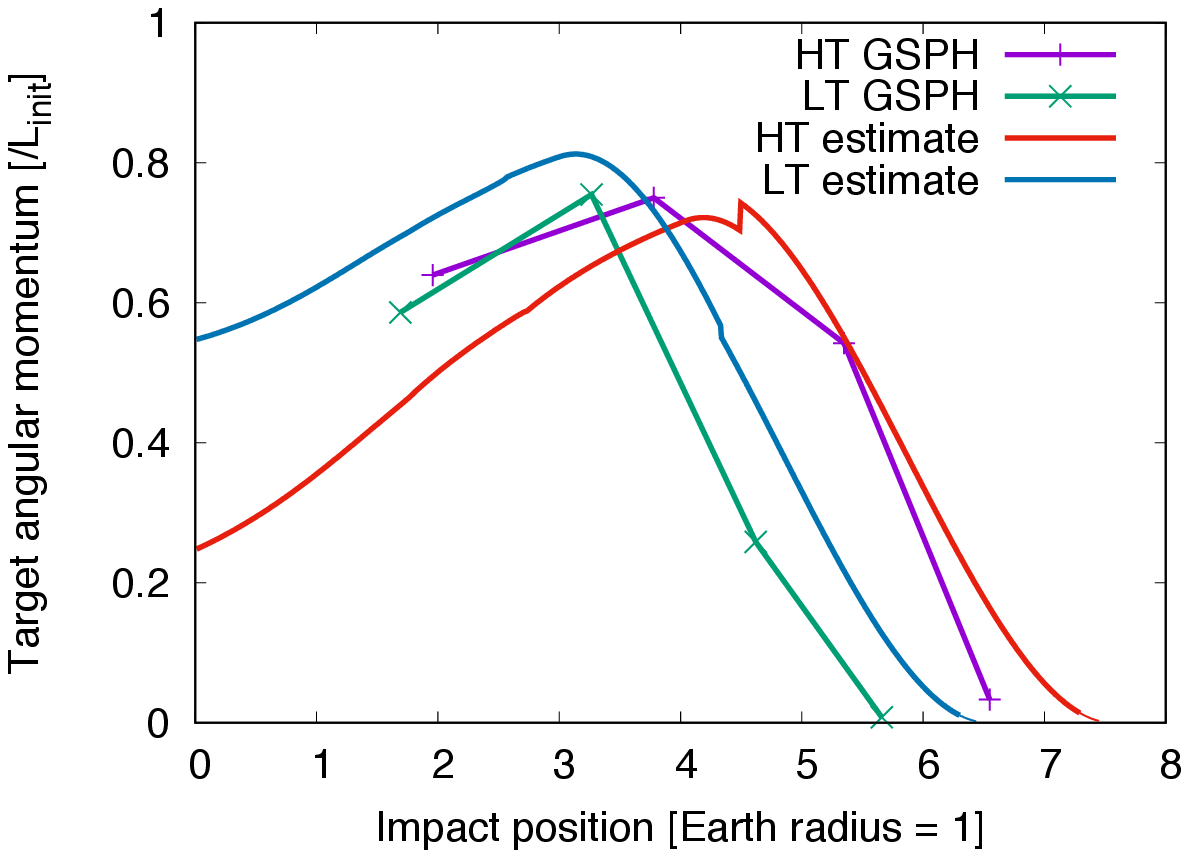}
\includegraphics[width=7cm]{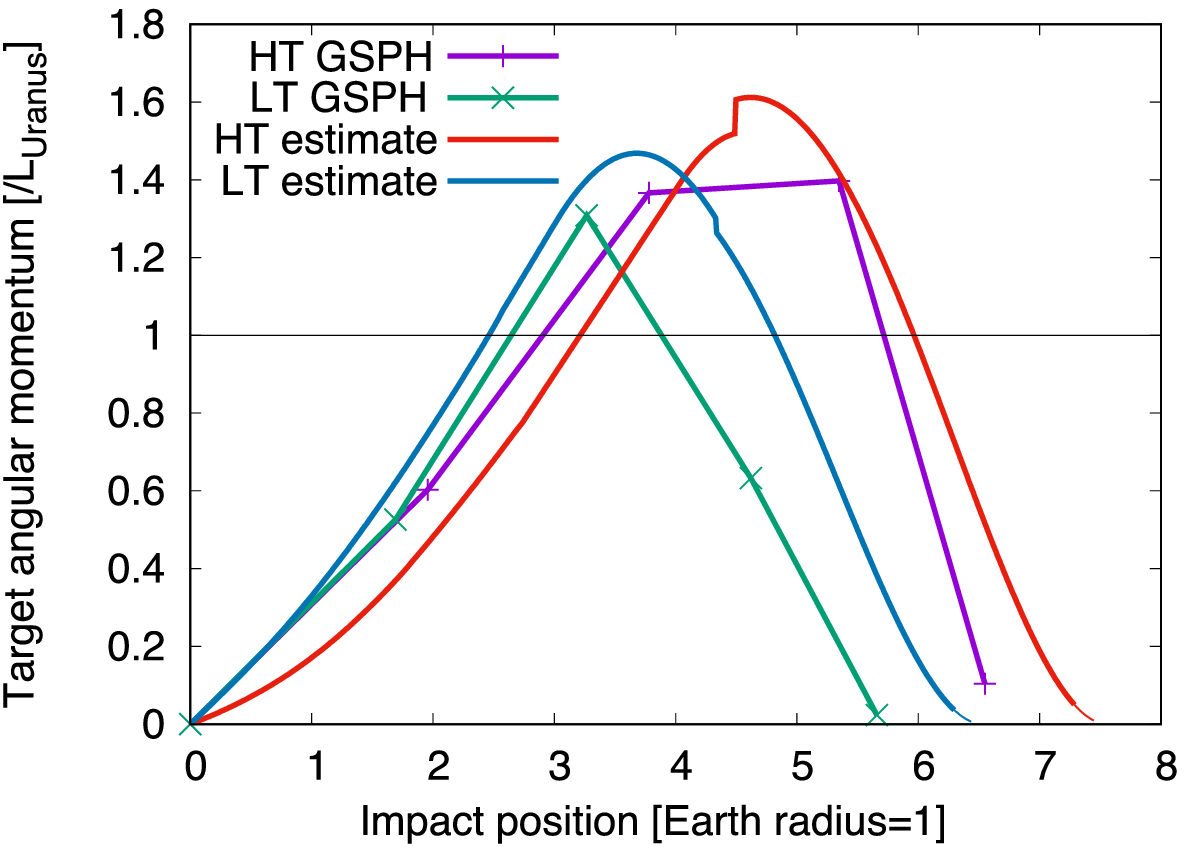}
\caption{\label{fig_mloss_Lz}
This figure shows the relationship between the obtained angular momentum and impact positions.
The purple and green line represents the high temperature target and low temperature target, respectively.
The Blue and orange lines represent the analytical solution for the high temperature target and low temperature target derived by Eq.~\ref{Lp_target}.
Left panel shows the efficiency of the transported angular momentum by a giant impact.
Right panel shows the obtained angular momentum normalized by present angular momentum of Uranus \citep{Podolak1987}.
}
\end{center}
\end{figure}

Figure~\ref{fig_mloss_Lz} shows the relationship between the obtained angular momentum and impact parameter.
It shows that the angular momentum gain of the target is related to the eroded mass.
The trend of the obtained angular momentum can be understood by Eq.~\ref{Lp_target}.
Moreover, the ejecta velocity is also understood by the ejecta scaling law (Eq.~\ref{vej_melosh}).
The right panel of figure~\ref{fig_mloss_Lz} shows the relationship for the obtained angular momentum normalized by present angular momentum of Uranus \citep{Podolak1987} and impact parameter.
Our result suggests that HT case explain the present angular momentum of Uranus even if the impactor's mass is $1~M_\oplus$.
If the target is in the high entropy state, the hydrogen envelope is expanded.
Then the cross-section of proto-Uranus is enlarged and the angular momentum transported by the impactor is larger than in the LT case.
We assumed that the initial hydrogen envelope is 20~\%, which is larger than present Uranus.
\citet{Venturini2016} implied that  low- and intermediate-mass planets (mini-Neptunes to Neptunes) can be formed with total mass fractions of hydrogen up to 30~\% considering the envelope polluted by ice materials. 
Thus, proto-Uranus might have had a more massive envelope than present Uranus
because the target can loose envelope gas as a result of a giant impact.

\subsection{Angular momentum transfer}
\citet{Hyodo2017_ring} showed that the tidal disruption of a passing Kuiper Belt object is able to explain the formation of current ring and inner regular satellites of Uranus.
On the other hand, previous studies \citep{Slattery1992,Kegerreis2018} and our simulation suggest that a giant impact is also able to supply material to form regular satellites of Uranus.
The total mass of major regular satellites around Uranus (Miranda, Ariel, Umbriel, Titania and Oberon) is $\sim 8.8\times10^{24}~$g \citep{Brown1991}, which is equivalent to $10^{-4}$ by Uranus's mass.
Moreover, 
the total angular momentum of them is $\sim 10^{-2} \times L_\mathrm{U}$.
In this section, we consider the condition for particles that are bounded gravitationally and not fall onto the target after the impact.
Those particles are supposed to be the origin of circumplanetary disk or ring.

\begin{figure}[h]
\begin{center}
\includegraphics[width=14cm, bb=10 10 400 400]{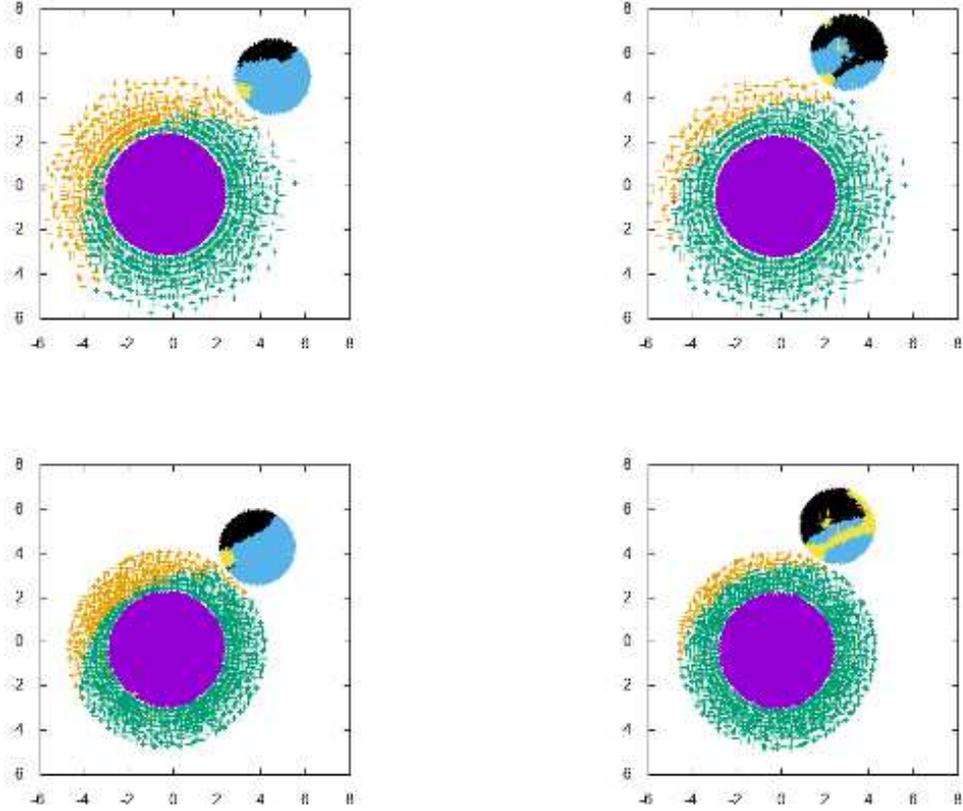}
\caption{\label{peri_region}
Particles that become the circum-planetary disk are shown with different colors in the initial condition.
Only the particles close to the midplane $(-0.1 < z < 0.1)$ are shown.
The color code is the same as Figures 2,3 except that in addition,
red and black dots shows hydrogen and water particle that satisfy Eq~\ref{cond1}, respectively.
In our simulation, most of the hydrogen particles which satisfy Eq~\ref{cond1} also satisfy Eq~\ref{lost_cond_1}. That is, those particle should eroded, which shown in orange dots.
Thus the number of red particles are very small in this representation.
Left-top and right-top figure shows the result of HT target whose impact angles are $45^\circ$ (left-top) and $60^\circ$ (right-top), respectively. 
Left-bottom and right-bottom figure show the result of LT target whose impact angles are $45^\circ$ (left-bottom) and $60^\circ$ (right-bottom), respectively. 
}
\end{center}
\end{figure}

Here we propose the criterion which determines the ejected particles will reaccrete or not by using the particle's orbital element.
That is, they will no reaccrete if the pericenter distance of the $i$-th particle is longer than the target radius;
\begin{equation}
a_i ( 1-e_i)  >  R_p \label{cond1}
\end{equation}
where $a_i$ and $e_i$ are semimajor axis and eccentricity of the $i$-th particle, respectively.
$a_i$ and $e_i$ are calculated by 
\begin{eqnarray}
a_i &=& \left( \frac{2}{|\mathbf{x}_i- \mathbf{x}_g|^2} - \frac{|\mathbf{v}_i-\mathbf{v}_g|^2}{G(M_p+m_i)}  \right)^{-1} \\
e_i &=& \sqrt{1-\frac{|(\mathbf{x}_i-\mathbf{x}_g)\times (\mathbf{v}_i-\mathbf{v}_g)|}{G(M_p+m_i)a_i} }
\end{eqnarray}
where $\mathbf{x}_g$ and $\mathbf{v}_g$ are the position and velocity of the center of the gravity, respectively \citep[cf.][]{SSD}.
Figure~\ref{peri_region} shows the particles which satisfy the condition of Eq~\ref{cond1}.
Our result shows that particles satisfying Eq~\ref{cond1} and being bounded gravitationally exist in the case of $\theta=45^\circ, 60^\circ$.
Such particles mainly come from the impactor and their compositions are ice.
Previous study \citep{Slattery1992, Kegerreis2018} also demonstrated the same conclusion.

\subsection{Implication for the satellite formation}

In this study, the impact velocity is fixed to the escape velocity of the target, 
which means that all particles are bounded in the system gravitationally. 
The hydrogen envelope is eroded due to the momentum exchange with impactor.
After that, the hydrogen envelope is scattered by the impact.
Particles gravitationally unbound will escape, while the others will accrete onto the target.

\begin{figure}[h]
\begin{center}
\includegraphics[width=7cm, bb=10 10 400 400]{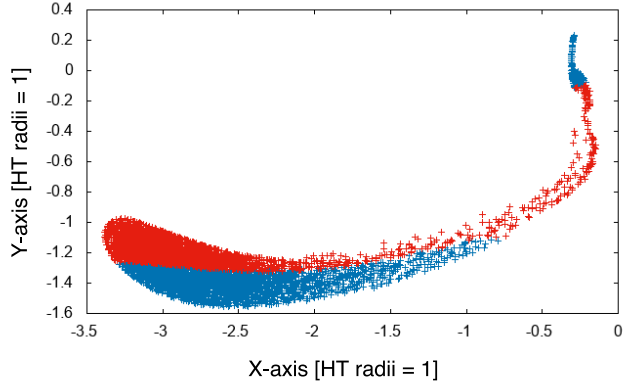}
\caption{\label{imp_Lz_ex} 
The transport of angular momentum via self-gravity.
This snapshot is taken $3~t_\mathrm{ff}$ after
the collision for the case of the impact angle $60^\circ$ on HT target.
The color shows the time derivative of the angular momentum due to the gravitational force.
Red and blue show $dL_z/dt>0$ and $dL_z/dt<0$, respectively.
}
\end{center}
\end{figure}

On the other hand,
some fraction of the impactor material and small amount of material blown out from the target do not accrete onto the target and remain in orbits around the target.
During the impact event,
hydrodynamical and tidal forces stretch the material,
and then the angular momenta are exchanged by self-gravitational and hydrodynamical force.
Figure~\ref{imp_Lz_ex} shows the time derivative of the particle's angular momentum.
Particles in the direction of moving transfer their angular momentum to particles behind them
and then the orbital distances of the latter increase.
If $dL_z/dt$ are positive, particles should move outward
because they gain angular momentum,
while particles whose $dL_z/dt$ are negative fall onto the target.
Consequently, 
materials from the impactor are left around the target.

\begin{figure}[h]
\begin{center}
\includegraphics[width=7cm]{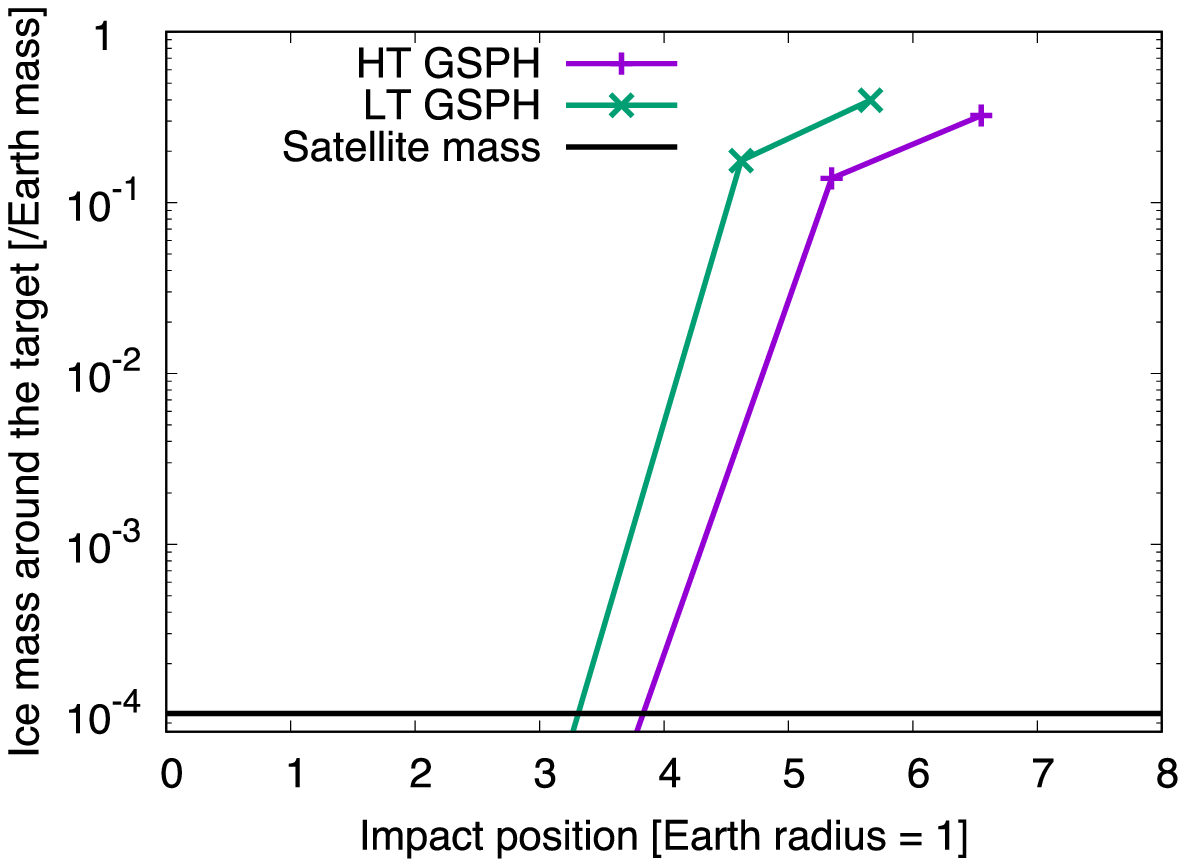}
\includegraphics[width=7cm]{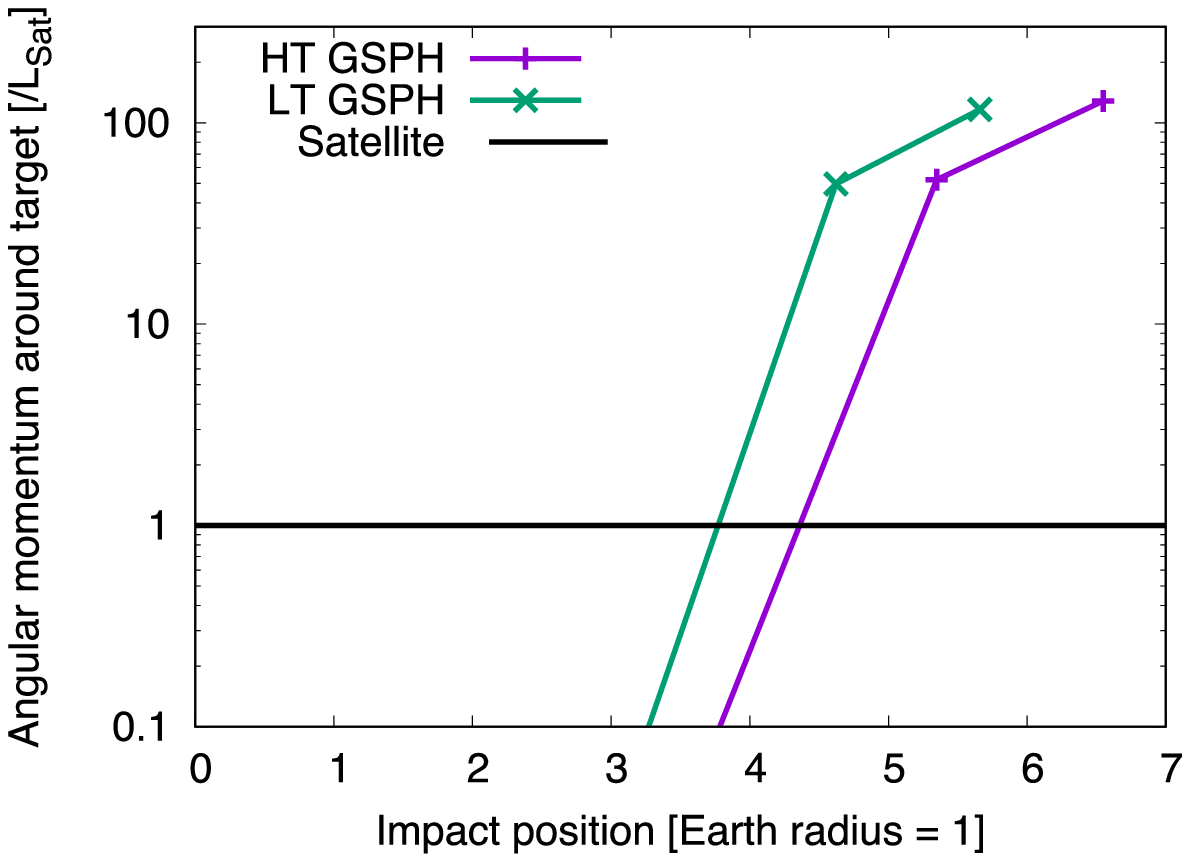}
\caption{\label{fig_mloss_ice_mat}
Relationship between the impact angle and total water particles mass (left panel) and their angular momentum (right panel)
in the circum-planetary disk.
Purple and red lines represent the high temperature target, while green and blue lines are low temperature target, respectively.
The black line shows the total mass and angular momentum of regular satellite of Uranus.
}
\end{center}
\end{figure}

Figure~\ref{fig_mloss_ice_mat} shows the particle mass which satisfy Eq.~\ref{cond1}.
Our simulation shows that an impact event
can supply a sufficient amount of ices for the formation of the regular satellites.
In the present simulation, we have adopted the impactor composed only of water, just for simplicity. In reality, the impactor should contain more or less rocky materials although the impactor's ice-to-rock ratio is unknown. We expect that rocky material should also be ejected depending on the impact parameters, if we perform a simulation with the impactor that contains rocky material. However, the determination of the ejected material in the case of a rocky impactor is beyond the scope of the first paper in this line of our work.
Comparing to \citet{Kegerreis2018}, the mass in orbit according to our result is a factor of two lower.
Our result also suggests that a LT target leads to a larger amount of material in orbit than a HT target does (Figure~\ref{fig_mloss_ice_mat}).
Since the entropies of our targets are higher than assumed in \citet{Kegerreis2018}, our results are consistent with that work.

\section{Conclusion} \label{conclusion}

In this paper we have performed numerical simulations
of a giant impact on a young Uranus-like ice giant
with Godunov SPH simulation with realistic structures composed of ice and hydrogen-helium gas in the case of no initial rotation of the target.
We find that there is a relationship between the resultant mass loss and the angular momentum of the target.
Our results suggest that a giant impact on proto-Uranus can
explain the present value of the angular momenta of Uranus and its satellite system.
We also find that 
if the target is in high entropy state, it obtains larger angular momentum and lose its envelope more efficiently than the low entropy state,
because the hydrogen envelope of the high entropy target is significantly more extended.
Our results also show that less mass remains gravitationally bound in high entropy target than in a low entropy target.
Our results may provide a step forward to understand the origin of Uranus.

\acknowledgments

KK thanks Y. Guo and K. Sugiura for discussions on GSPH method.
He also thanks N. Hosono and D. Namekata for the use of FDPS.
The authors also thank K. Iwasaki, S. Takasao, and H. Kobayashi for fruitful comments. 
Numerical computations were carried out on Cray XC30 at Center for Computational Astrophysics, National Astronomical Observatory of Japan.
This work was supported by JSPS KAKENHI Grant Numbers 16H02160.

\appendix

\section{Riemann solver for Godunov SPH with self-gravity}
In this appendix, we introduce the Riemann solver for our GSPH \citep[see also][]{Guo}.
Here we consider the Riemann solver for $i$ and $j$ particles.
The expression for the result of Riemann problem we use is 
\begin{eqnarray}
p^{\ast} &=& \frac{p_i/W_i+p_j/W_j -v_i+v_j}{1/W_i+1/W_j} \label{past1} \\
v^{\ast} &=& \frac{v_i W_i+v_j W_j -p_i+p_j}{W_i+W_j} \label{vast1}
\end{eqnarray}
where $p, v$ are pressure, and velocity, respectively.
$W_k$ abbreviates
\begin{eqnarray}
W_k &=& \sqrt{\gamma p_k\rho_k} \sqrt{1+\frac{\gamma+1}{2\gamma}\frac{p^\ast-p_k}{p_k}}~~\mathrm{for}~p^\ast \ge p_k \\
W_k &=& \sqrt{\gamma p_k\rho_k} \frac{\gamma-1}{2\gamma}\frac{1-p^\ast/p_k}{1-(p^\ast/p_k)^{(\gamma-1)/(2\gamma)}}~~\mathrm{for}~p^\ast < p_k 
\end{eqnarray}
where $\gamma, \rho$ are the ratio of specific heats and density, respectively.
To take into account the gravitational force $\mathbf{F}_g$ on Eq.~\ref{past1} and \ref{vast1},
we replace $p_i$ and $p_j$ with $p'_i$ and $p'_j$ as shown in the following:
\begin{eqnarray}
p'_i &=& p_i - \frac{1}{2} \rho_i C_{s,i} \mathbf{F}_g\cdot \hat{s}\delta t \\
p'_j &=& p_i + \frac{1}{2} \rho_j C_{s,j} \mathbf{F}_g\cdot \hat{s}\delta t
\end{eqnarray}
where $C_{s,k} = \sqrt{\gamma_k p_k \rho_k}$, $\hat{s}=(\mathbf{x}_i-\mathbf{x}_j)/|\mathbf{x}_i-\mathbf{x}_j|$ and $\delta t$ is the time step.

\section{Thermal evolution of the target}
\begin{figure}[h]
\begin{center}
\includegraphics[width=7cm]{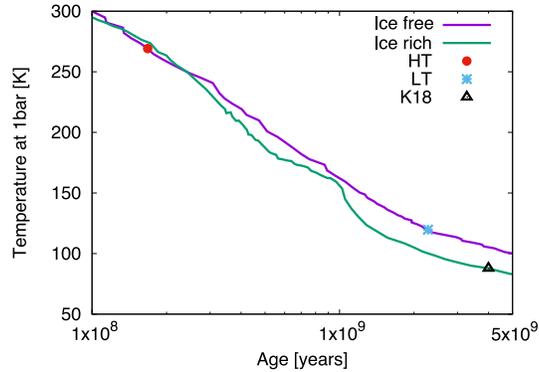}
\caption{\label{evolv_T1bar}
Relationship between the age and the temperature at 1~bar.
Purple and green lines represent thermal evolution of ice giants whose ice mass fractions in the atmosphere are 0~wt\% and 50~wt\% (ice rich), respectively.
The red and blue means HT and LT target conditions, respectively.
The condition for \citet{Kegerreis2018} is represented black (K18 in the figure), which corresponds to the present Uranus.
}
\end{center}
\end{figure}

In this appendix, we briefly explain the thermal evolution of the target.
The target is composed of 80~\% of water core surrounded by 20~\% of hydrogen-helium atmosphere.
We numerically integrate the thermal evolution. We assume the target consists of three layers in spherical symmetry and hydrostatic equilibrium, that include, from top to bottom (1) a radiative-convective equilibrium atmosphere composed of hydrogen helium, (2) a convective equilibrium envelope composed of hydrogen helium, and (3) a convective equilibrium water-ice core.
Details of atmosphere, interior, and thermal evolution model are described in \citet{Kurosaki2017}.
Figure~\ref{evolv_T1bar} shows the thermal evolution of the target.
Purple line shows an ice giant whose hydrogen-helium atmosphere is free from ice materials,
while green line shows the case with atmosphere is mixed with ice materials by 50 \% by mass.
The temperature at 1~bar of present Uranus is $\sim 80$~K.
Temperatures at 1~bar for HT and LT target are 270~K and 120~K, respectively.
HT and LT correspond to $1.6\times10^8$ years and $2.3 \times10^9$ years, respectively. 
As previous studies \citep{Fortney2011,Nettelmann2016} indicated that Uranus could imply strong barrier to interior convective cooling.
\citet{Kurosaki2017} suggested that an ice-rich atmosphere can accelerate the cooling due to the effect of the condensation in the atmosphere, though it requires 50 \% of ice by mass.
Our target assume an ice giant with ice-free atmosphere. 
In our model, the target in \citet{Kegerreis2018} may corresponds to an evolved ice giant with ice-rich atmosphere.
Therefore, we are proposing that
if HT target's interior is mixed with ice efficiently after the impact,
subsequent thermal evolution of HT can reproduce present Uranus.
However, the detailed determination of the amount of mixing by the impact is beyond the scope in this study.

\section{The determination of unbound particles: the effect of the internal energy}
\begin{figure}[h]
\begin{center}
\includegraphics[width=7cm]{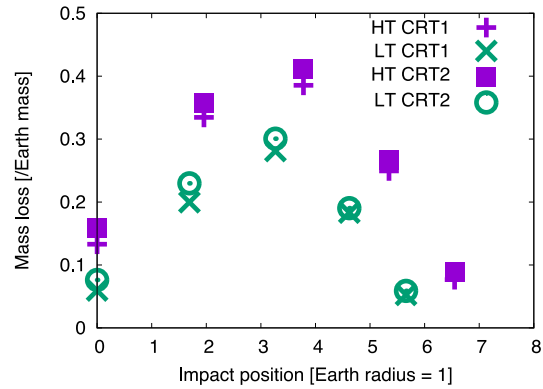}
\caption{\label{mloss_crit_hikaku}
Relationship between the eroded mass and the impact parameter.
Purple and green symbols represent the high temperature target and low temperature target using criterion CRT1 (squares, circles) or CRT2 (crosses, pluses).
}
\end{center}
\end{figure}

We check the effect of the internal energy on the determination of unbound particles.
The condition that the particle's velocity exceeds the escape velocity (hereafter CRT1) is  Equation~\ref{lost_cond_1}.
On the other hand, 
the particle's internal energy increase after the impact event. 
The larger internal energy causes the larger pressure to the particle.
Thus the pressure gradient may accelerate the particle and contribute to its escape.
We include the internal energy into the condition, Equation~\ref{lost_cond_1}; 
\begin{eqnarray}
m_i u_i + \frac{1}{2}m_i |\mathbf{v}_i|^2 - m_i \sum_{i\ne j}\frac{Gm_j}{|\mathbf{r}_i - \mathbf{r}_j|}  &>& 0 \label{lost_cond_1c},
\end{eqnarray}
and hereafter we call this condition, CRT2.
Figure~\ref{mloss_crit_hikaku} shows the relationship between the eroded mass and the impact parameter.
We find that the eroded mass of CRT2 is larger than that of CRT1. 
When the impact occurs, the particle's internal energy is increased
because the impact event distributes the kinetic energy to the internal energy of the particles.
After the impact,
the hydrodynamic motion redistributes
the particle's internal energy to the kinetic energy \citep[cf.][]{Genda2015}.
That is, the pressure gradient caused by the internal energy accelerates the particles.
However, the difference of the eroded mass between CRT1 and CRT2 is less than 5~\%.
Therefore, we adopt CRT1 as the condition of unbound particles.

\end{document}